\documentclass{article}
\usepackage{graphicx} % Required for inserting images
\usepackage[a4paper, total={7in, 10in}]{geometry}
\PassOptionsToPackage{table,x11names}{xcolor}
\usepackage{caption}
\usepackage{subcaption}
\usepackage{comment}
\usepackage{amsmath}
\usepackage{amssymb}
\usepackage{array}
\usepackage{xcolor}
\usepackage{multicol}
\usepackage{nomencl}
\makenomenclature
\usepackage{lmodern}
\usepackage{float}
\usepackage{paralist}
\usepackage{authblk}
\usepackage{wrapfig}
\usepackage{soul}
\usepackage{cases}

\usepackage[backend=biber,doi=false,isbn=false,url=false,eprint=false]{biblatex} 

\title{1D solutions for compressible two-phase flows in a heated and cooled duct: mechanical equilibrium}

\author[1]{S. Schropff\thanks{solene.schropff@univ-amu.fr}}
\author[1]{F. Petitpas\thanks{fabien.petitpas@univ-amu.fr}}
\author[1]{E. Daniel\thanks{eric.daniel@univ-amu.fr}}
\affil[1]{Aix Marseille Univ, CNRS, IUSTI, Marseille, France}

\begin{document}

\include{figures}

\maketitle

\begin{abstract}
    Analytical/quasi-analytical solutions are proposed for a steady, compressible, two-phase flow in mechanical equilibrium in a rectilinear duct subjected to heating followed by cooling. The flow is driven by the pressure ratio between a variable outlet pressure and an upstream tank. A critical pressure ratio distinguishes subsonic and supersonic outlet regimes: the article proposes a methodology to determine the full flow behaviour, as a function of pressure ratio and heat-flux distribution. Going forward, these analytical reference solutions will help validate numerical codes for more complex industrial applications. Specific results are studied for a mixture of liquid water and water vapour.
\end{abstract}

\section{Introduction}

Analytical solutions for heated compressible flows have firstly been developed for simple 1D cases \cite{shapiro1953dynamics}. To obtain those solutions, the conditions of the flow must be specified at the inlet. However, when the flow is subsonic, acoustic waves will travel up and down the duct, resulting in a change of inlet state. Thus, to consider upstream variations and determine solutions, the inlet is connected to a tank, therefore providing constant inlet stagnation state. Consequently, subsonic flows can be recovered without specifying the flow quantities.

In doing so, the study provided in \cite{fathalli2018} has shown that, at most, a heated flow in a duct connected to a tank can accelerate up until the sonic point. Then, further research presented in \cite{schropff2023} described what happens once a cooled subsection is appended: if the end of the heated subsection reaches the sonic state, then the cooled subsection can either bring the flow back to the subsonic state, or it can accelerate it into a supersonic regime. This branching depends on the pressure ratio between the outlet and the tank: according to this pressure ratio and the applied function of heat flux, the behaviour of the flow can be determined. As the tank conditions are given fixed data, variations in flow behaviour can be obtained by varying the outlet pressure.

This previous study has defined the possible regimes and critical pressure ratios; it was demonstrated that there was no possible steady shock wave, as the flow is ruled by the intersection of its \emph{Rayleigh line} and \emph{Crussard curve}. This description shows a similarity with the study of flows in nozzles. The role of the area of the cross section of the nozzle was correlated to the local amount of heat applied on the duct: however, it was shown that where a single ratio of areas is needed to determine nozzle flows, both heating and cooling power values are needed for heated and cooled flows.

In the following paper, we study a two-phase flow under mechanical equilibrium: both phases evolve freely in terms of thermal and chemical aspects but share a single velocity and single pressure. This kind of flow was first presented in \cite{kapila2000two} for condensed granular matter. It then has been largely used to describe liquid-gas mixtures \cite{murrone2005} \cite{saurel2009}. The prime and novel objective of this study is to propose analytical reference solutions for this type of flow, following the previously described set-up \cite{schropff2023}: the solutions are extended to a one-dimensional two-phase steady compressible flow, subjected to heating, and then cooling power. These reference solutions are developed using the “Stiffened-Gas” equation of state \cite{harlow1968}, allowing to consider both liquids and gases. This work will help validate numerical tools for the simulation of non-adiabatic compressible flows, but also to size engineering processes and installations.

\section{Two-phase flow model: mechanical equilibrium}

\subsection{Unsteady model}
\label{sec:model}

The following system describes the baseline unsteady model of our study, for a two-phase, compressible, inviscid flow in a mechanical equilibrium. It is an extension of the model presented in \cite{kapila2000two}, with the addition of an external heat source term $\delta \dot q$ [$\text{W.m}^{-3}$]:

\begin{equation}
    \label{eqn:system}
    \left\{
    \begin{array}{l}
        \partial_t \alpha_1 + \textbf{u}.\nabla.\alpha_1 = K \nabla.\textbf{u} + \Theta \delta \dot q / \rho \\
        \partial_t (\alpha_1 \rho_1) + \nabla.(\alpha_1 \rho_1 \textbf{u}) = 0 \\
        \partial_t (\alpha_2 \rho_2) + \nabla.(\alpha_2 \rho_2 \textbf{u}) = 0 \\
        \partial_t \rho + \nabla.(\rho \textbf{u}) = 0 \\
        \partial_t (\rho \textbf{u}) + \nabla.(\rho \textbf{u} \otimes \textbf{u} + P \textbf{I}) = 0 \\
        \partial_t (\rho E) + \nabla.((\rho E + P) \textbf{u}) = \delta \dot q
    \end{array}
    \right.
\end{equation}

Both phases share a single pressure $P$ and velocity \textbf{u}. Mixture quantities are defined as followed: $v=1/\rho=\sum_kY_k v_k$, $E=e+0.5 u^2$, $e=\sum_k Y_k e_k$, where $v$ is the mixture specific volume, $\rho$ is the mixture density, $E$ is the mixture total energy and $e$ is the mixture specific internal energy.

Each symbol indexed by $(\cdot)_k$ denotes a phase $k$ variable: $v_k$, $\rho_k=1/v_k$ and $e_k$ are respectively the specific volume, density, and internal energy of said phase. $Y_k$ represents its mass fraction within the mixture, whereas $\alpha_k=Y_k v_k/v$ is its volume fraction. It must be noted that $\sum_k Y_k =\sum_k \alpha_k =1$. The mixture entropy is defined as $s=\sum_k Y_k s_k $ and the entropy equation for phase $k$ is the following, with $s_k$ the entropy and $T_k$ the temperature of the phase:

\begin{equation}
\label{eqn:entropy_k}
    \rho \frac{ds_k}{dt} = \frac{\delta \dot q}{T_k}    
\end{equation}

With $c_k^2= \partial P_k/\partial \rho_k )_{s_k}$ the definition for the speed of sound of phase $k$, the mixture sound speed of this model is the Wood sound speed $c_w$ \cite{wood1956textbook}:

\begin{equation}
    \label{eqn:c_wood}
    \frac{1}{\rho c_w^2} = \sum_k \frac{\alpha_k}{\rho_k c_k^2}
\end{equation}

The compressibility factor for a two-phase flow is $K=(\rho_2 c_2^2 - \rho_1 c_1^2 )/(\rho_1 c_1^2/\alpha_1 + \rho_2 c_2^2/\alpha_2)$: $K \nabla.\textbf{u}$ describes the variations of the volume fraction under acoustic perturbations. We demonstrate $\Theta=(\rho_2 \Gamma_2 - \rho_1 \Gamma_1 )/(\rho_1 c_1^2/\alpha_1 + \rho_2 c_2^2/\alpha_2)$, where $\Gamma_k=v_k \partial P_k/ \partial_k )_{\rho_k}$ is the Gruneisen coefficient of phase $k$. For a positive heat source term, if phase $1$ is denser than phase $2$ (therefore $\rho_1>\rho_2$), the volume fraction $\alpha_1$ will decrease.

The system is closed using a convex equation of state (EOS) $e_k=e_k (P_k,v_k)$ for each phase $k$. In this paper, the “Stiffened-Gas” (SG) EOS is used: it allows to describe both liquids and gases as the equation considers attractive and repulsive effects in matter:

\begin{equation}
    \label{eqn:energy_EOS}
    e_k(P_k, v_k) = \frac{P_k + \gamma_k P_{k,\infty}}{\gamma_k - 1} v_k + e_{k,ref}
\end{equation}

The EOS parameters $\gamma_k$, $P_{k,\infty}$ and $e_{k,ref}$ are obtained from reference thermodynamic curves, characteristics of the material and transformation under study (see \cite{lemetayer2004} for details). Therefore, by using SG EOS, the speed of sound of phase $k$ reads:

\begin{equation}
    \label{eqn:celerity_phase}
    c_k^2 = \gamma_k (P_k + P_{k,\infty}) v_k
\end{equation}

%%%%%%%%%%%%%%

\subsection{Steady-state analysis}

\subsubsection{System description}

Let us consider a one-dimensional steady flow in a rectilinear duct, defined by a length $L$, a constant cross-section $S$ and a circumference $C$. The outer surface $S_{ext}=C\times L$ of the duct receives a 2-steps function of heat flux $\varphi(x)$  [$\text{W.m}^{-2} $] along axis x ; $x_{heat}$ denotes the end of the heated subsection and the beginning of the cooled subsection (defining the subscript $(\cdot)_{heat}$). This yields a corresponding power function defined as $Q(x)=\int_0^x \varphi(\eta)  \,C\, d\eta$ [W]:

$$
\varphi(x) = \left\{
    \begin{array}{ll}
        \varphi_{heat} > 0 & \mbox{if } x \leq x_{heat} \\
        \varphi_{cool} < 0 & \mbox{otherwise}
    \end{array}
\right. \Longrightarrow
    Q(x) = \left\{
    \begin{array}{ll}
        C \, x \, \varphi_{heat} & \mbox{if } x \leq x_{heat} \\
        C \left[ x_{heat} \, \varphi_{heat} + (x-x_{heat}) \, \varphi_{cool} \right] & \mbox{otherwise}
    \end{array}
\right.
$$

Therefore, the power received by the entire heated subsection is $Q_{heat} = C \, x_{heat} \, \varphi_{heat}$, the power lost by the entire cooled subsection is $Q_{cool}=C \, (L-x_{heat} ) \, \varphi_{cool}$ and the total power received by the duct is $Q_{out} = Q_{heat} + Q_{cool}$. Finally, the power received per unit area of cross-section defined previously can be retrieved at any point $x$ by the following relationship: $\dot q_{s,x} = Q(x)/S \varpropto Q(x)$.

%%%%%%%

\begin{figure*}[ht]
    \centering
    \includegraphics[width=0.8\textwidth]{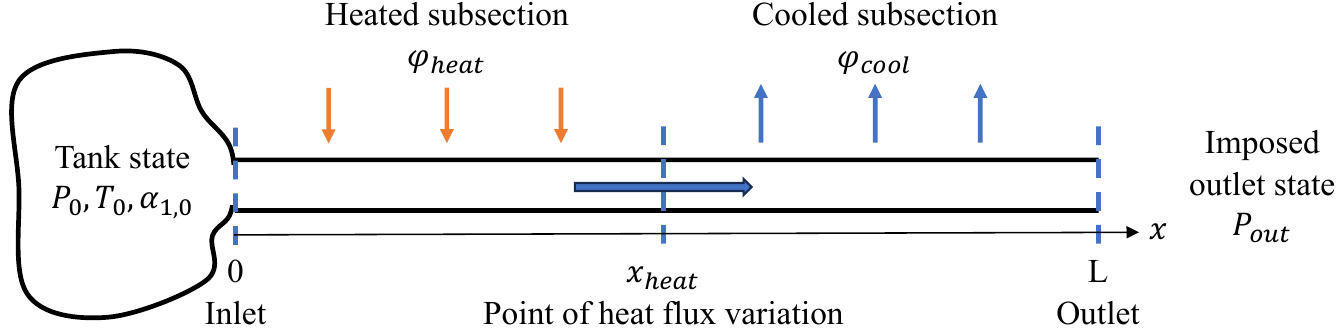}
    \caption{Fluid flowing from a tank into a heated and cooled duct.}
    \label{fig:system}
\end{figure*}

The inlet of the duct is connected to a tank, which is characterized by tank conditions with zero velocity, known stagnation pressure, temperature, and phases distribution. By coupling both systems, the flow is prescribed along the duct based on given parameters: stagnation conditions, outlet pressure, and an applied heat flux function. The system is depicted in Figure~\ref{fig:system}.

%%%%%%%%%%

\subsubsection{Duct relations}

Integration of system~(\ref{eqn:system}) at steady-state between two points $(\cdot)_{in}$ and $(\cdot)_x$ along axis $x$ of the duct yields:

\begin{equation}
    \label{eqn:mass}
    \rho_{in}u_{in} = \rho_{x} u_{x} = \dot m_{s} %\equiv \frac{\dot m}{S}
\end{equation}

\begin{equation}
    \label{eqn:qdm}
    \rho_{in}u_{in}^2 + P_{in} = \rho_{x}u_{x}^2 + P_{x}
\end{equation}

\begin{equation}
    \label{eqn:nrj}
    \dot m_{s}(H_{x} - H_{in}) = \dot q_{s,x}
\end{equation}

where $\dot m_{s} = \rho u = \dot m/S$ is the mass flow rate per unit area of cross-section, $\dot q_{s} = Q/S$ is the power received per unit area of cross-section and $H=h+0.5 u^2$ is the total specific enthalpy of the mixture. The mixture internal enthalpy $h$ is defined as $h=\sum_k Y_k h_k $, with $h_k=e_k+P_k v_k$ the internal specific enthalpy of phase $k$. Combining equations (\ref{eqn:mass}) and (\ref{eqn:qdm}) yields the equation of the \textit{Rayleigh line} of the mixture in the $(P,v)$ plane:

\begin{equation}
    \label{eqn:rayleigh}
    P_{x} = \dot m_{s}^2(v_{in} - v_{x}) + P_{in}
\end{equation}

As we hypothesize a uniform distribution of heat flux between the phases, we add the following relation between $(\cdot)_{in}$ and $(\cdot)_{x}$ to close the problem:

\begin{equation}
\label{eqn:crussard_energy}
    e_{k,x} - e_{k,in} + \frac{P_{x} + P_{in}}{2}(v_{k,x} - v_{k,in}) = \frac{\dot q_{s,x}}{\dot m_s} 
\end{equation}

This guarantees the conservation of the mixture energy:

\begin{equation}
\label{eqn:crussard_totalenergy}
    \sum_k Y_k e_{k,x} - \sum_k Y_k e_{k,in} + \frac{P_{x} + P_{in}}{2} \left( \sum_k Y_k v_{k,x} - \sum_k Y_k v_{k,in} \right) = \sum_k Y_k \frac{\dot q_{s,x}}{\dot m_s} 
\end{equation}

Combining equation (\ref{eqn:crussard_energy}) with the SG EOS expression of phase internal energy $e_k (P_k,v_k )$ (\ref{eqn:energy_EOS}) yields the expression of phase \textit{k} specific volume, which describes the \emph{Crussard curve} for phase \textit{k} in the \emph{(P,v)} plane:

\begin{equation}
\label{eqn:vk_crussard}
    v_{k,x}(P_{in}) = \frac{ (\gamma_k+1)(P_{in} + P_{k,\infty}) + (\gamma_k-1)(P_{x} + P_{k,\infty}) }{ (\gamma_k+1)(P_{x} + P_{k,\infty}) + (\gamma_k-1)(P_{in} + P_{k,\infty}) } v_{k,in} + \frac{ 2(\gamma_k-1)  }{ (\gamma_k+1)(P_{x} + P_{\infty}) + (\gamma_k-1)(P_{in} + P_{k,\infty}) } \frac{\dot q_{s,x}}{\dot m_s}
\end{equation}

\subsubsection{Tank-inlet relations}

The tank, denoted by subscript $(\cdot)_0$, contains a mixture at mechanical and thermal equilibrium. The two phases are respectively in $Y_{k,0}=\alpha_{k,0} \rho_{k,0}/ \rho_0$ mass proportions, which are conserved from the tank throughout the duct in mechanical equilibrium as there is no phase change, yielding $Y_{k,0} \equiv Y_k$. The phase $k$ density within the tank is retrieved through the SG EOS, as $\rho_{k,0}=\rho_{k,0}(P_0,T_0)$. The inlet (denoted by subscript $(\cdot)_{in}$) stagnation state is recovered by slowing down the flow to rest through an adiabatic process and is equivalent to the tank state. This means that $P_{0,in} \equiv P_0$ and $T_{0,in} \equiv T_0$. 

On the tank-inlet side, the only available relations are the conservation of total specific enthalpy ($H_{in}=H_0$) and specific entropy ($s_{in}=s_0$) of the mixture. Even if the supersonic solution is mathematically acceptable, only the subsonic solution is sought to be consistent with an actual tank. Therefore, the mass flow rate is expressed as followed:

\begin{equation}
\label{eqn:massflowrate}
    \dot m_s(P_{in}) = \frac{\sqrt{2(h_0 - h_{in})}}{v_{in} }
\end{equation}

As there is no possible shock wave or input of heat between the tank and the inlet, the model is also isentropic from the point of view of each phase $k$ ($ds_k/dt=0$). With the use of the SG EOS, we obtain the following equations:

\begin{equation}
    \label{eqn:volume_inlet_phase}
    v_{k,in}^{is}(P_{in}) = v_{k,0} \left(\frac{P_0 + P_{k,\infty}}{P_{in} + P_{k,\infty}}\right)^{1/\gamma_k}
\end{equation}

\begin{equation}
    \label{eqn:enthalpy_inlet_phase}
    h_{k,in}^{is}(P_{in}) = \frac{\gamma_k v_{k,0} (P_0 + P_{k,\infty})}{\gamma_k - 1} \left(\frac{P_0 + P_{k,\infty}}{P_{in} + P_{k,\infty}}\right)^{(1-\gamma_k)/\gamma_k} + e_{k,ref}
\end{equation}

\section{Reference solutions: determining flow behaviour}

In the following analysis, the characterization of the flow is done from the point of view of the \textbf{mixture}. Two-phase flows behaviour can be divided in specific regimes by the same critical pressure ratios as for single-phase flows subjected to heating and cooling powers \cite{schropff2023}: $\Pi_1$ and $\Pi_3$ (see Table~\ref{tab:Rp}). The flow can be subsonic, supersonic, or choked (sonic) at specific locations. We find out by analytical calculations what are the conditions required to determine the flow regime and we seek the values of the different critical pressure ratios by solving the system of equations composed of the duct relations and the tank-inlet relations.

We define the dynamical parameter of the flow, which is the pressure ratio between the outlet and the tank: $\Pi=P_{out}/P_0$. As $P_0$ is a fixed parameter, the outlet pressure $P_{out}$ will be modified to cause a change in behaviour. 

\begin{center}
    \begin{tabular} {| m{0.18\textwidth} | m{0.75\textwidth} |}
        \hline
        $\Pi = 1$ & No flow \\
        \hline
        $\Pi_{1} < \Pi < 1$ & Fully subsonic flow \\
        \hline
        $\Pi = \Pi_{1}$ & Subsonic inlet, choked flow at the end of the heated subsection and subsonic everywhere \\
        \hline
        $\Pi_3 < \Pi < \Pi_{1}$ & Subsonic inlet, choked flow at the end of the heated subsection and supersonic outlet (shock waves outside of the duct) \\ 
        \hline
        $\Pi = \Pi_{3}$ & Subsonic inlet, choked flow at the end of the heated subsection and supersonic outlet (adapted flow outside of the duct) \\
        \hline
        $\Pi < \Pi_{3}$ & Subsonic inlet, choked flow at the end of the heated subsection and supersonic outlet (expansion waves outside of the duct) \\ 
        \hline
        
    \end{tabular}
    \captionof{table}{Pressure ratios definition for a flow in heated and cooled duct.}
    \label{tab:Rp}
\end{center}

%%%%%%%

\subsection{Fully subsonic flow}

When the mixture is fully subsonic between the inlet and the outlet, it means that $\Pi > \Pi_1$: information coming from downstream can travel back up and impact the inlet state. Given the previous statement, to firstly determine the inlet mixture pressure $P_{in}$ and therefore the inlet state, we evaluate the \textit{Rayleigh line} (\ref{eqn:rayleigh}) between the inlet $(\cdot)_{in}$ and outlet $(\cdot)_{out}$:

\begin{equation}
\label{eqn:fPin_subsonic}
    f(P_{in}) = \dot m_s^2 \left( \sum_k Y_k v_{k,out} - \sum_k Y_k v_{k,in}^{is} \right) + P_{out} - P_{in} = 0
\end{equation}

On one hand, the isentropic relations allow to determine the mass flow rate $\dot m_s(P_{in})$ (\ref{eqn:massflowrate}) and inlet specific volumes of the phases $v_{k,in}^{is}(P_{in})$ (\ref{eqn:volume_inlet_phase}). On the other hand, the outlet pressure $P_{out}$ is a known constant, as well as the outlet heat flux $\dot q_{s,out}$, and thus the specific phase volumes at the outlet, $v_{k,out}(P_{in})$, are computed from equation (\ref{eqn:vk_crussard}).

Now that the inlet state is known, the flow state along each point of the duct must be determined. Knowing the inlet pressure $P_{in}$, the unknown is now the duct pressure at any point \textit{x}, $P_x$. To evaluate it, we consider again the \textit{Rayleigh line} (\ref{eqn:rayleigh}), but this time between the known inlet and undetermined point \textit{x} of the duct:

\begin{equation}
\label{eqn:fPx_subsonic}
    f(P_{x}) = \dot m_s^2 \left( \sum_k Y_k v_{k,x} - \sum_k Y_k v_{k,in}^{is} \right) + P_{x} - P_{in} = 0
\end{equation}

All inlet related variables, $P_{in}$, $\dot m_s(P_{in})$ and $v_{k,in}^{is}(P_{in})$ are known and constant for a given pressure ratio $\Pi$. Therefore, the remaining variable to determine are the specific phase volumes $v_{k,x}(P_x)$ from equation (\ref{eqn:vk_crussard}), given that we know the heat flux function $\dot q_{s,x}$ at all points of the duct. Equation (\ref{eqn:fPx_subsonic}) yields two positive roots: the lowest one corresponds to a supersonic regime, which is not our case of study, and the highest one corresponds to a subsonic regime, which is therefore the one that is retained for the value of subsonic pressure $P_x^{sub}(P_{in})$.

%%%%%%%%%

\subsection{Choked flow}
\label{sec:choked_flow}

We now consider that the mixture is choked at $x_{heat} \equiv (\cdot)_{heat}$ point: it means that $\Pi \leq \Pi_1$ and information coming from the cooled subsection cannot impact the heated subsection and the inlet state. Having reached a choked flow at the end of the heated subsection, we assume that $M=1$ at that point, and that there are two possible states in the cooled subsection: choked and then supersonic, or subsonic. Therefore, we now evaluate the \textit{Rayleigh line} (\ref{eqn:rayleigh}) between $(\cdot)_{in}^*$ and $(\cdot)_{heat}$, to numerically determine $P_{in}^*$, where $(\cdot)_{in}^*$ denotes the inlet state for which the flow ends up choked:

\begin{equation}
\label{eqn:fPin_choked}
    f(P_{in}^*) = \dot m_s^2 \left( \sum_k Y_k v_{k,heat} - \sum_k Y_k v_{k,in}^{is} \right) + P_{heat} - P_{in}^* = 0
\end{equation}

The specific phase volumes at the inlet $v_{k,in}^{is}(P_{in}^*)$ are determined from equation (\ref{eqn:volume_inlet_phase}) and the mass flow rate $\dot m_s(P_{in}^*)$ from equation (\ref{eqn:massflowrate}). The remaining variable, $P_{heat}$, which is the mixture pressure at the end of the heated subsection, will further constrain the system to ensure a sonic state at that point. The conservation of total energy (\ref{eqn:nrj}) between $(\cdot)_{in}^*$ and $(\cdot)_{heat}$ yields the following equation (given that $H_in=H_0=\sum_k Y_{k,0} h_{k,0}$):

\begin{equation}
\label{eqn:fPheat}
    f(P_{heat}) = H_{heat} - H_0 - \dot q_{s,heat}/\dot m_s = 0
\end{equation}

Determining $P_{in}^*$ thus requires imbricating two numerical methods to solve the equations presented in (\ref{eqn:fPin_choked}) and (\ref{eqn:fPheat}). The total mixture enthalpy at $(\cdot)_{heat}$ point for a choked flow ($M=1 \rightarrow u=c$) is expressed as followed: 

\begin{equation}
\label{eqn:Hheat_choked}
    H_{heat}(P_{heat}) = \sum_k Y_k h_{k,heat} + 0.5 c_{heat}^2
\end{equation}

The Wood sound speed $c_w$ is evaluated by equation (\ref{eqn:c_wood}) at point $(\cdot)_{heat}$. The phase specific internal enthalpy $h_k = e_k + P_k \, v_k$ is determined from the SG EOS (\ref{eqn:energy_EOS}).

Once the inlet state has been determined for a choked flow, we want to determine the flow state along the entirety of the duct. With the same approeach as for the fully subsonic flow, equation~(\ref{eqn:rayleigh}) is evaluated between $(\cdot)_{in}^*$ and $(\cdot)_{x}$:

\begin{equation}
\label{eqn:fPx_choked}
    f(P_{x}) = \dot m_s^2 \left( \sum_k Y_k v_{k,x} - \sum_k Y_k v_{k,in}^{is} \right) + P_{x} - P_{in}^* = 0
\end{equation}

For the heated subsection, between $(\cdot)_{in}^*$ and $(\cdot)_{heat}$, the flow can only be subsonic. Therefore, the subsonic solution is determined by retaining the highest positive root of equation (\ref{eqn:fPx_choked}), $P_x^{sub} (P_{in}^*)$. For the cooled subsection ($x\geq x_{heat}$), the flow can either be subsonic — then again $P_x = P_x^{sub}(P_{in}^*)$ — or supersonic, where $P_x = P_x^{sup}(P_{in}^*)$, which is the lowest positive root of equation (\ref{eqn:fPx_choked}). The critical outlet states are determined for $(\cdot)_{out}$: therefore $\Pi_1=P_{out}^{sub}/P_0$ and $\Pi_3=P_{out}^{sup}/P_0$.

\section{Results: variation of pressure ratio}

We now present some results that illustrate the flow behaviour in various conditions. They are obtained by solving the analytical solutions proposed above, in the following configuration:

\begin{compactitem}
    \item Water vapour $(\cdot)_v$: $\gamma_v=1.358$, $C_{v,v}=1247$, $e_{ref,v}=1.97 \times 10^6$ [J.kg$^{-1}$], $P_{v,\infty}=0$ [Pa] (particular case of SG EOS)
    \item Liquid water: $\gamma_l=3.423$, $C_{v,l}=1231.2$, $e_{ref,l}=-1.15 \times 10^6$ [J.kg$^{-1}$], $P_{l,\infty}=8.99 \times 10^8$ [Pa]
    \item Tank state: $P_0 = 2$ [Bar], $T_0 = 394$ [K], $\alpha_{0,l}=0.99$, $\alpha_{0,v}=0.01$
    \item Geometry (cylinder of radius $R$ and volume $V$): $V=1/4\pi$ [$m^3$], $R=1/2\pi$ [m], $S=\pi R^2 = 1/4\pi$ [$m^2$], $C=2 \pi R = 1$ [m], $L=1$ [m], $x_{heat}=L/2$ [m]
    \item Power: $Q_{heat}=100$ kW, $Q_{cool}=-100$ kW ($Q_{out}=0$) 
\end{compactitem}

The critical pressure ratios can first be calculated by the method presented in Section \ref{sec:choked_flow}, which yields the following values: $\Pi_1=0.0828$, $\Pi_3=0.0762$. To depict the multiple possible behaviours that are described in Tab.~\ref{tab:Rp}, the pressure ratio $\Pi$ must be specified and can then be varied. The whole solution is then calculated between the inlet and the outlet.

\begin{figure*}[h!]
\centering
    \begin{subfigure}[h]{0.49\textwidth}
        \includegraphics[width=\textwidth]{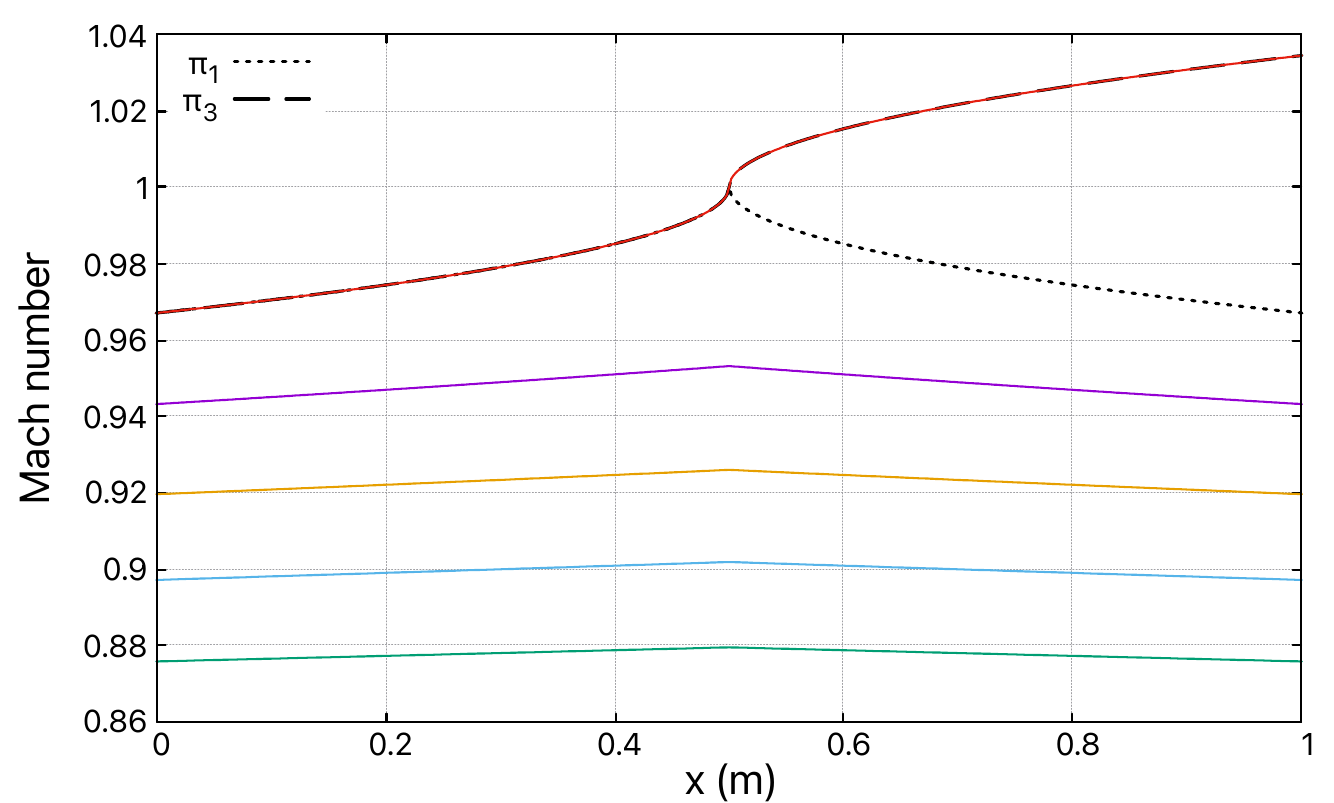}
        \label{fig:varRp_M_mix}
    \end{subfigure}
    \begin{subfigure}[h]{0.49\textwidth}
        \includegraphics[width=\textwidth]{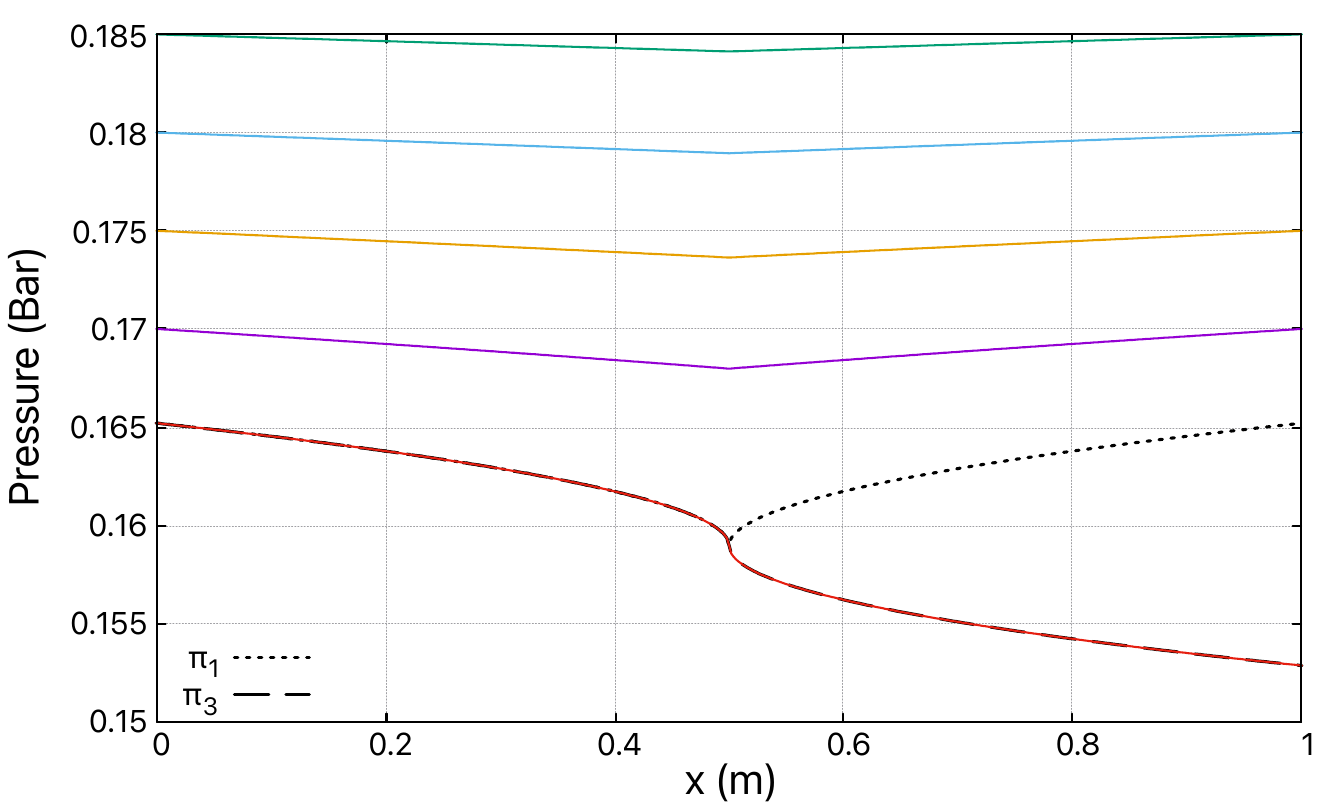}
        \label{fig:varRp_P_mix}
    \end{subfigure}
    \caption{Behaviour of heated and cooled mixture of water and vapour under mechanical equilibrium, for different values of pressure ratio (different colours).}
    \label{fig:varRp_mix}
\end{figure*}

\begin{figure*}[h!]
\centering
    \begin{subfigure}[h]{0.49\textwidth}
        \includegraphics[width=\textwidth]{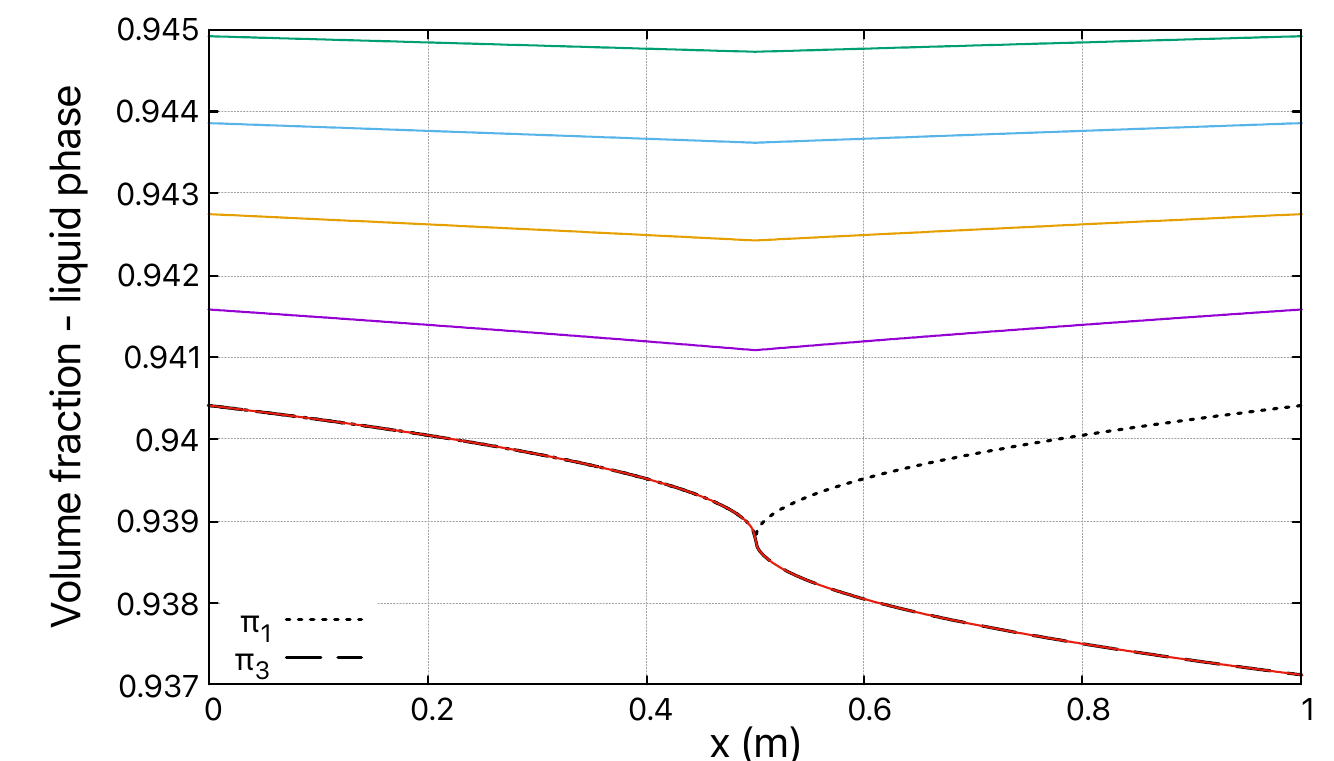}
        \label{fig:varRp_a_liq}
    \end{subfigure}
    \begin{subfigure}[h]{0.49\textwidth}
        \includegraphics[width=\textwidth]{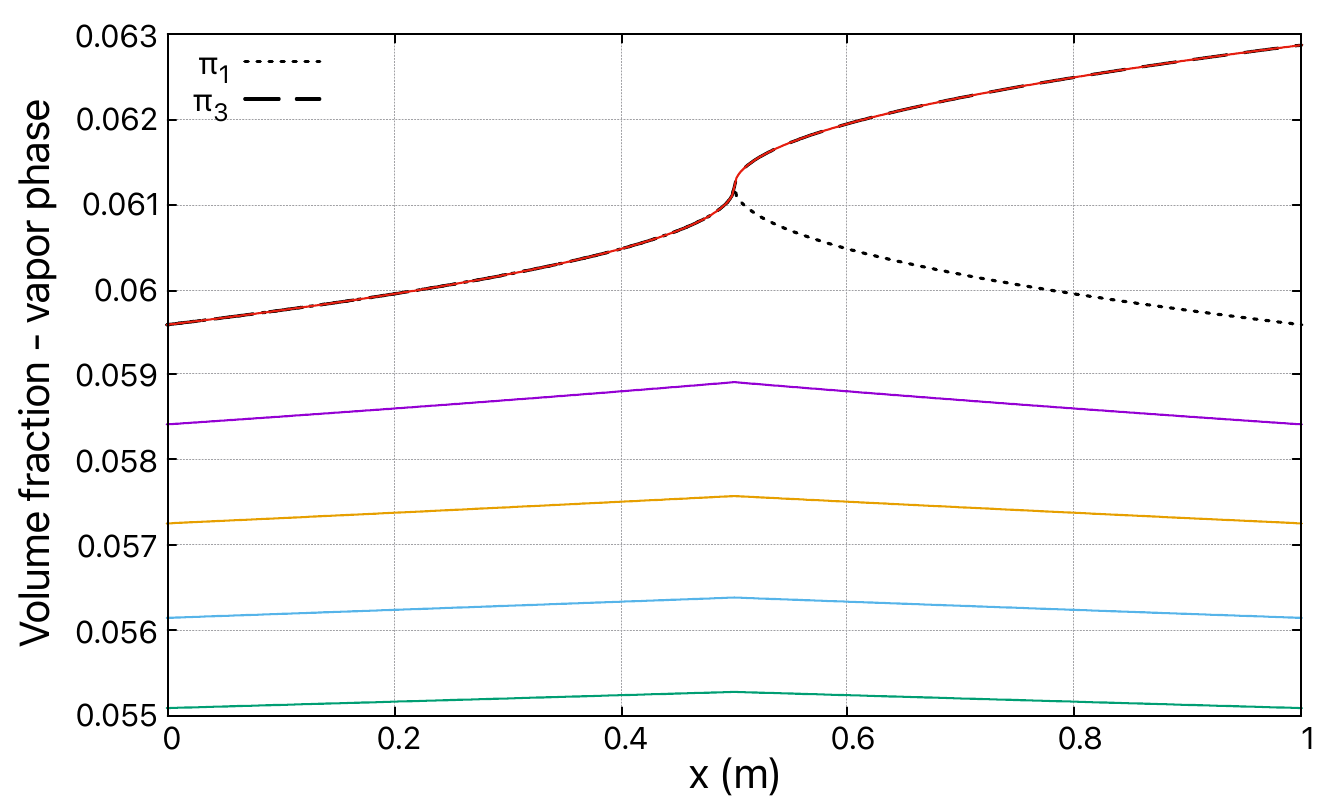}
        \label{fig:varRp_a_vap}
    \end{subfigure}
    %%%%
    \begin{subfigure}[h]{0.49\textwidth}
        \includegraphics[width=\textwidth]{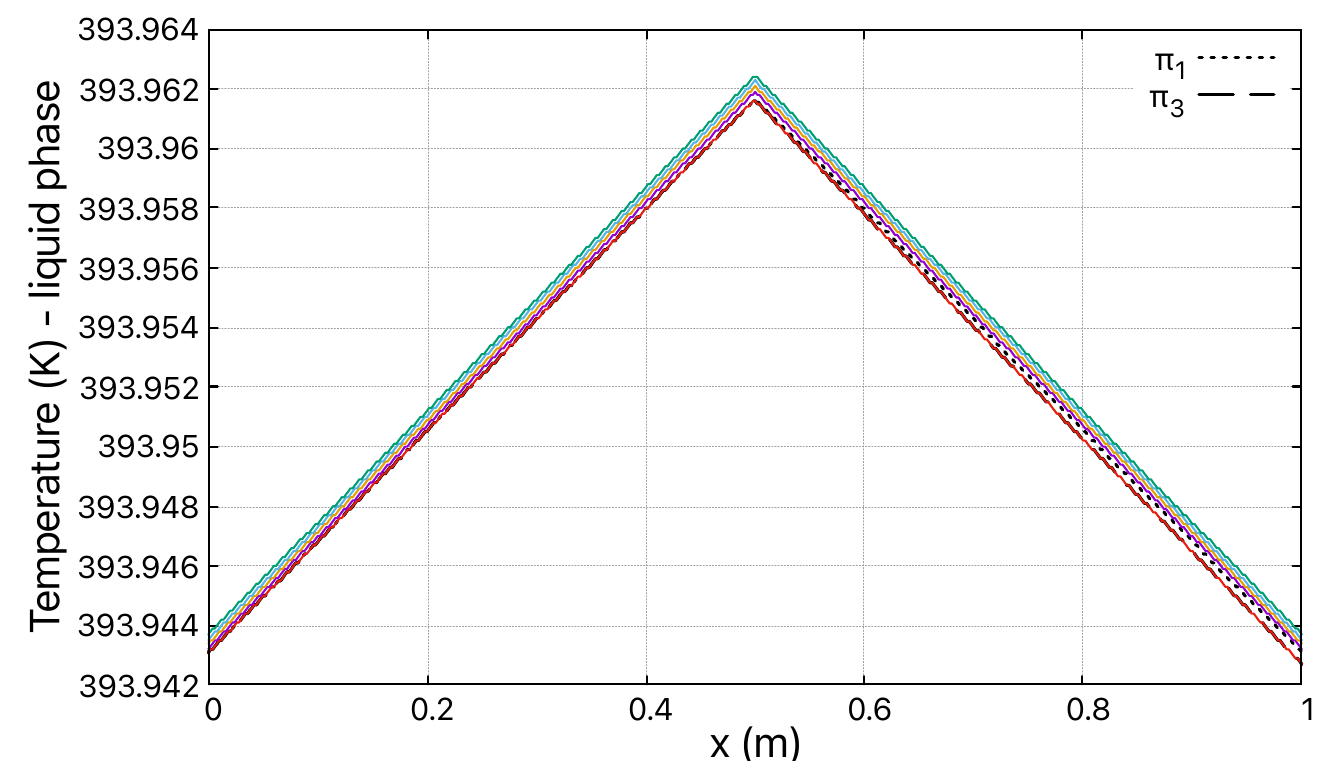}
        \label{fig:varRp_T_liq}
    \end{subfigure}
    %%%%
    \begin{subfigure}[h]{0.49\textwidth}
        \includegraphics[width=\textwidth]{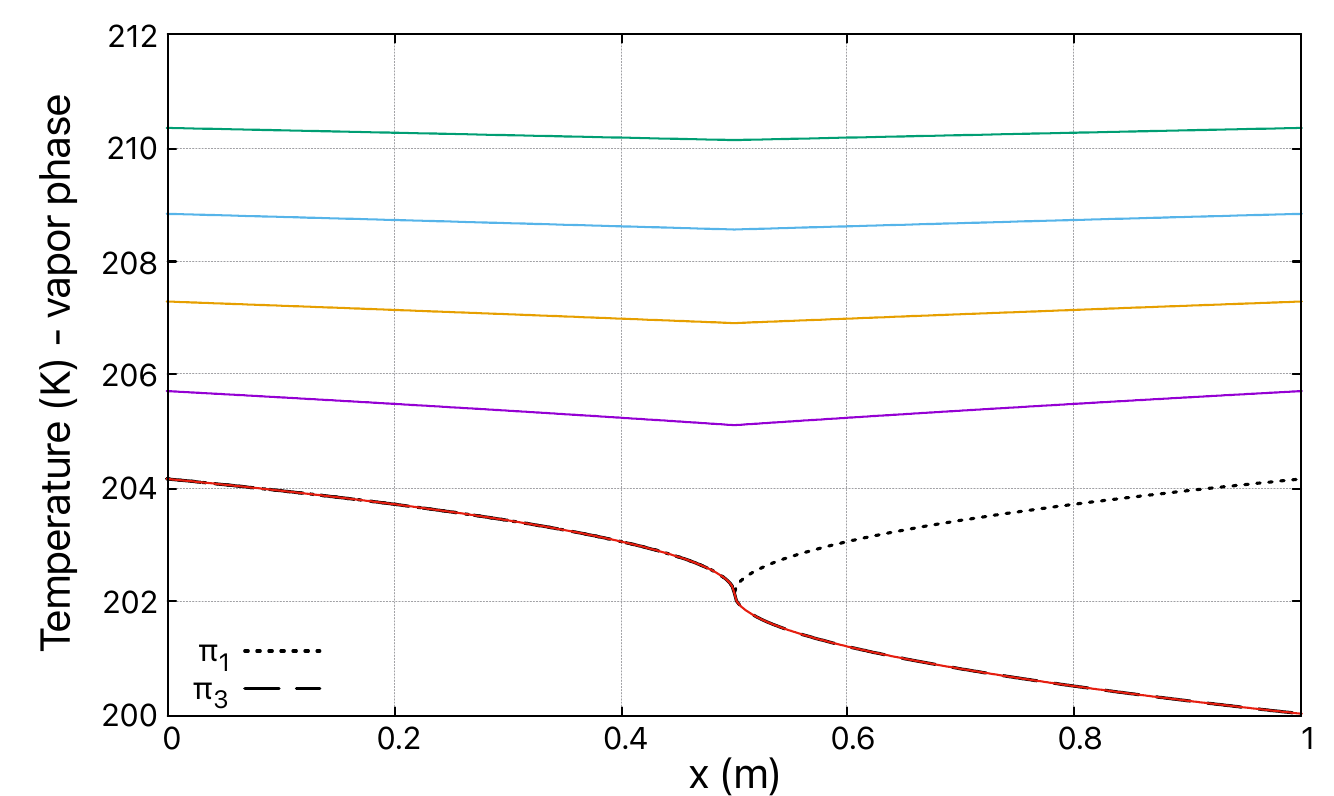}
        \label{fig:varRp_T_vap}
    \end{subfigure}
    \caption{Behaviour of heated and cooled liquid water (left) and water vapour (right) under mechanical equilibrium, for different values of pressure ratio (different colours). }
    \label{fig:varRp_phases}
\end{figure*}

Fig.~\ref{fig:varRp_mix} displays mixture variables such as pressure and Mach number. Fig.~\ref{fig:varRp_phases} displays phase variables for liquid water and water vapour, such as volume fraction and temperature. All these variables are represented for various values of $\Pi \in [0.0825; 0.0925]$, themselves depicted in different colours. The specific solutions corresponding to $\Pi_1$ and $\Pi_3$ are labelled in the figures, respectively as dashed and dotted lines.

%%%%%%%%%%%%%%

\subsection{Analysis of results for $\Pi>\Pi_1$}

Multiple solutions are represented for $\Pi>\Pi_1$: the mass flow rate $\dot m_s (P_{in} )$ varies as the state of the flow in the heated subsection still depends on the variations of $P_{out}$. When $\Pi$ decreases but the flow is still subsonic, the flow overall accelerates (increase of Mach number) and its pressure decreases. Locally, the Mach number increases in the heated subsection, and then decreases in the cooled subsection, which follows the evolution described in \cite{shapiro1953dynamics}.

The volume fraction of water decreases when the flow is heated and increases when it is cooled. This is related to the description provided in Section~\ref{sec:model}: heating the denser phase (here, the water) contracts it. Now considering their temperature, the evolution is not so straightforward. For the water, the decrease of $\Pi$ only causes an overall slight decrease of temperature, and generally, the temperature follows an obvious pattern: it increases when the flow is heated and decreases when it is cooled. The temperature of the vapour, however, behaves counter-intuitively: it decreases in the heated part and increases in the cooled part. For single-phase flows, this specific evolution happens when $1/\sqrt{\gamma}<M<1$ \cite{shapiro1953dynamics}.

%%%%%%%%%%%%%%

\subsection{Analysis of results for $\Pi \leq \Pi_1$}

Once $\Pi \leq \Pi_1$, the mixture is choked at the end of the heated subsection ($M_{heat}=1$). It can remain subsonic in the cooled part if $\Pi=\Pi_1$, or become supersonic if $\Pi < \Pi_1$: this branching only depends on the applied pressure ratio $\Pi$. When $\Pi \leq \Pi_1$, the mass flow rate $\dot m_s (P_{in}^* )$ is fixed by heated subsection choked state, which is determined and cannot change. If $\Pi < \Pi_1$, then whatever the value of $\Pi$, every supersonic solution is overlapping (see red curves) and there is no steady shock wave in the cooled subsection. 

The difference in range of variation for each phase is noticeable: while the solutions for the vapour are quite distinct and display a clear change of behaviour in the cooled subsection once $\Pi < \Pi_1$, the solutions for the water are only slightly varying between different pressure ratios. This may be caused by the gap between the critical pressure ratios of the mixture and the critical pressure ratios of the single-phase flow, which is presented in \cite{schropff2023}.

\section{Conclusion}

Following the studies done on solely heated \cite{fathalli2018} and then on heated and cooled single-phase compressible flows \cite{schropff2023} connected to a tank, reference solutions were developed for two-phase compressible heated and cooled flows in a mechanical equilibrium.

Starting from the unsteady model, a steady-state analysis was developed to underline the relations defining the flow within a non-adiabatic duct. Then, by coupling the inlet of said duct to a tank with known fixed conditions, the state of the flow was prescribed throughout the duct. The definition of two possible states was provided: either the pressure ratio between the outlet of the duct and the tank is greater than a specific critical pressure ratio and the mixture remains fully subsonic, or the pressure ratio is lower and therefore the cooled subsection triggers a supersonic flow. The method to determine this critical pressure ratio was developed, depending on the value of the heat flux function applied to the duct. Results and analysis were then provided for a mixture of liquid water and vapour.

The analytical solutions stemming from this work are expected to be used for the validation of non-adiabatic compressible flows numerical simulation tools, which are largely used in industrial applications (nuclear, aerospace), as those involve strong heat fluxes and thermal exchanges. Future works will consist in applying the same set-up on limit cases for other equilibrium models, such as thermal and thermodynamical equilibriums, which were studied in \cite{fathalli2018} for solely heated flows.

\printbibliography

\end{document}